\documentclass[twocolumn,prl,superscriptaddress,aps,showpacs,preprintnumbers,amsmath,amssymb,a4paper]{revtex4}
\usepackage{graphicx}
\usepackage{dcolumn}
\usepackage{bm}
\usepackage[english]{babel}
\usepackage{xcolor}
\begin{document}
\preprint{APS/123-QED}
\title{Avoided ferromagnetic quantum critical point: Unusual short-range ordered state in CeFePO}
\author{S.~Lausberg}
\email{lausberg@cpfs.mpg.de}
\affiliation{Max-Planck-Institute for Chemical Physics of Solids, D-01187 Dresden, Germany}
\author{J.~Spehling}
\affiliation{Institute for Solid State Physics, TU Dresden, D-01069 Dresden, Germany}
\author{A.~Steppke}
\affiliation{Max-Planck-Institute for Chemical Physics of Solids, D-01187 Dresden, Germany}
\author{A.~Jesche}
\altaffiliation[Present adress: ]{Ames Laboratory, US DOE and Department of Physics and Astronomy, Iowa State University, Ames, IA 50011, USA}
\affiliation{Max-Planck-Institute for Chemical Physics of Solids, D-01187 Dresden, Germany}
\author{H.~Luetkens}
\author{A.~Amato}
\author{C.~Baines}
\affiliation{Laboratory for Muon-Spin Spectroscopy, Paul-Scherrer-Institute, CH-5232 Villigen, Switzerland}
\author{C.~Krellner}
\altaffiliation[Present adress: ]{Institute of Physics, Goethe University Frankfurt, D-60438 Frankfurt am Main, Germany}
\affiliation{Max-Planck-Institute for Chemical Physics of Solids, D-01187 Dresden, Germany}
\author{M.~Brando}
\author{C.~Geibel}
\affiliation{Max-Planck-Institute for Chemical Physics of Solids, D-01187 Dresden, Germany}
\author{H.-H.~Klauss}
\affiliation{Institute for Solid State Physics, TU Dresden, D-01069 Dresden, Germany}
\author{F.~Steglich}
\affiliation{Max-Planck-Institute for Chemical Physics of Solids, D-01187 Dresden, Germany}
\date{\today}
\begin{abstract}
Cerium 4$f$ electronic spin dynamics in single crystals of the
heavy-fermion system CeFePO is studied by means of
ac-susceptibility, specific heat and muon-spin relaxation ($\mu$SR).
Short-range static magnetism occurs below the freezing
temperature $T_{g}\approx0.7$\,K, which prevents the system from
accessing the putative ferromagnetic quantum critical point. In the
$\mu$SR, the sample-averaged muon asymmetry function is dominated by
strongly inhomogeneous spin fluctuations below $10\,$K and exhibits a
characteristic time-field scaling relation expected from glassy spin
dynamics, strongly evidencing cooperative and critical spin
fluctuations. The overall behavior can be ascribed neither to
canonical spin glasses nor other disorder-driven mechanisms.
\end{abstract}
\maketitle
A long-standing question in the field of quantum criticality is
whether a ferromagnetic (FM) quantum critical point (QCP) generally
exists and, if not, which are the possible ground states of matter
that replace it. Quantum critical points occur when a material is
continuously tuned with an external parameter (pressure, magnetic
field, etc.) between competing ground states at zero
temperature~\cite{Sachdev1999,Lohneysen2007}. An FM-QCP then exists
when it is possible to shift the Curie transition temperature
$T_{C}$ of a ferromagnet continuously to zero where a second order
quantum phase transition takes place. Quantum phase transitions
occur at zero entropy and are driven by quantum rather than thermal
fluctuations. These fluctuations diverge at the QCP modifying the
excitation spectrum of a metal and leading to a fundamental
instability of Landau's Fermi liquid (FL)~\cite{Baym2004}. Typical
signatures of such a behavior are observed in magnetic, thermal and
transport properties and are referred to as non-Fermi-liquid (NFL)
phenomena~\cite{Stewart2001}.

Although there is clear evidence for the existence of
antiferromagnetic (AFM) QCPs, the FM-QCP case is controversial. In
recent years, substantial experimental and theoretical efforts were
made to further investigate this problem. However, a wide range of possibilities
exists. On theoretical grounds, a 3-dimensional (3D) FM-QCP is
believed to be inherently unstable, either towards a first order
phase transition or towards an inhomogeneous magnetic phase
(modulated/textured
structures)~\cite{Kirkpatrick2012,Chubukov2004,Conduit2009}.
Similar results have been obtained in
2D~\cite{Kirkpatrick2012,Maslov2006,Conduit2010}. Several clean (stoichiometric)
magnetic transition-metal compounds, like
MnSi~\cite{Pfleiderer2007} or
ZrZn$_{2}$~\cite{Uhlarz2004}, show NFL behavior close to a
FM instability, but the transition changes into a
first order one.
In other systems
the existence of a FM-QCP has been proposed, most notably in
Nb$_{1-y}$Fe$_{2+y}$~\cite{Brando2008},
Zr$_{1-x}$Nb$_{x}$Zn$_{2}$~\cite{Sokolov2006} or
SrCo$_{2}$(Ge$_{1-x}$P$_{x}$)$_{2}$~\cite{Jia2011} where the
FM-QCP is attained by chemical substitution.
However, in these cases the influence of disorder remains ambiguous.

More appropriate candidates for the study of FM-QCPs are U-, Yb or
Ce-based \textit{f}-electron metals~\cite{Stewart2001,Lohneysen2007},
since in these materials the NFL signatures are much more pronounced
due to their heavy-fermion (HF) character. However, while there is
quite a number of U-based systems showing either a first order FM
transition (UGe$_2$, UCoAl, UCoGe)~\cite{Aoki2012} or indications
for a FM QCP (UCu$_{5-x}$Pd$_{x}$~\cite{Bernal1995},
URh$_{1-x}$Ru$_{x}$Ge,~\cite{Huy2007}), the number of Yb-based
systems close to a FM QCP is very limited
(YbNi$_{4}$P$_{2}$~\cite{Krellner2011},
YbCu$_{2}$Si$_{2}$~\cite{Fernandez-Panella2011}).
Several systems, like CeRu$_2$Ge$_2$~\cite{Sullow1999} or CeRuPO~\cite{Macovei2010} where the FM transition temperature
is suppressed to $T=0$ by hydrostatic pressure, exhibit a change into AFM order before reaching the QCP.
There are Ce-based alloys (CePd$_{1-x}$Rh$_{x}$~\cite{Westerkamp2009}) and also $d$-electron
metals (Ni$_{1-x}$V$_{x}$~\cite{Ubaid2010}) where it seems that
local disorder-driven mechanisms such as Kondo disorder or the
quantum Griffiths phase (QGP) scenario are responsible for the NFL
properties~\cite{Miranda1997,CastroNeto1998,Vojta2006}. Broad and
strongly $T$ dependent NMR and $\mu$SR linewidths are indicative for
such disorder-driven mechanisms. As a consequence spin-glass-like
behavior is often found, e.g., in CePd$_{1-x}$Rh$_{x}$, and
power-law corrections to the thermodynamic and transport properties
as well as in the local spin dynamics are expected in a broad region
across the putative QCP. The global phase transition then becomes
smeared~\cite{Westerkamp2009,Vojta2003,Hoyos2008}.

In this context, the layered Kondo-lattice system CeFePO is a unique
candidate for studying FM-QCPs, since it is a clean non-magnetic
(non-superconducting) HF metal located very close to a FM-QCP with
strong FM fluctuations~\cite{Bruening2008,Zocco2011}. CeFePO is a
homologue of the quaternary iron pnictides. It evolves from a
long-range ordered FM ground state, when a small amount of arsenic
is substituted for phosphorus~\cite{Jesche2011,Jesche2012,Luo2010}. Less As
concentration leads to a continuous decrease of $T_{C}$ culminating
into a putative FM-QCP. The isovalent As substitution not only
introduces a volume effect (shortening mostly the $c$-axis) but also
increases locally the hybridization strength between the trivalent
Ce-4$f$ and the Fe-$3d$ conduction electrons leading to an
enhancement of the Kondo temperature, which is
approximately $10\,$K for CeFePO~\cite{Holder2010}. Below this temperature, the
susceptibility and the Knight shift become field dependent and the
NMR line width broadens. This identifies the onset of short-range FM
correlations essentially within the basal plane, evidencing
a strong anisotropy~\cite{Bruening2008}, which could also be confirmed by recent NMR measurements
on oriented powder~\cite{Kitagawa2011}.
The ground state of CeFePO was found to be a paramagnetic heavy FL~\cite{Bruening2008}.

\begin{figure}[t]
\begin{center}
\includegraphics[width=0.95\columnwidth]{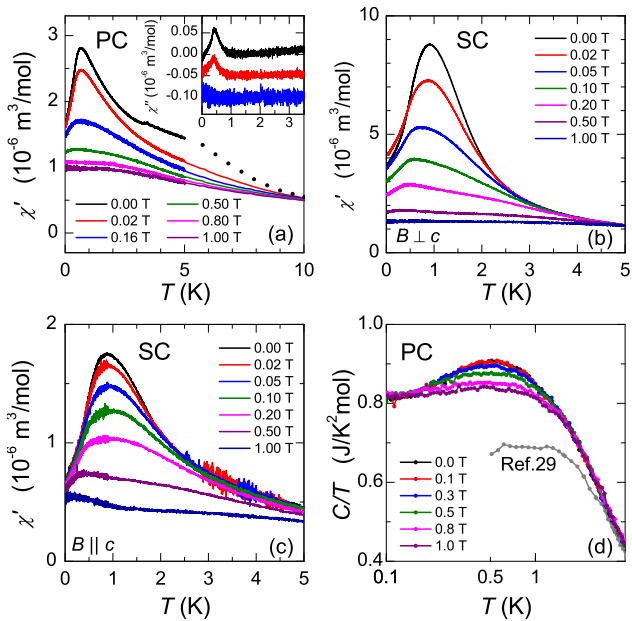}
\end{center}
\caption[1]{(Color online) (a) Ac-susceptibility $\chi^\prime(T)$ of
the PC sample with its imaginary part $\chi^{\prime\prime}(T)$
(inset).
(b) $\chi^\prime(T)$ vs $T$ for the SC sample with $B \perp c$.
(c) $\chi^\prime(T)$ vs $T$ for the SC sample with $B \| c$.
(d) Specific heat of the PC sample plotted as $C/T$ vs $T$.
}
\label{fig1}
\end{figure}

In this letter, we present a comprehensive study of the
ac-susceptibility ($\chi'$), specific heat ($C$) and muon-spin
relaxation ($\mu$SR) measurements on recently grown high-quality
single crystals of CeFePO. We find evidence of strongly inhomogeneous
spin fluctuations starting below $10\,$K and of a spin-glass-like
freezing at $T_{g} \approx 0.7$\,K which prevents the system from
accessing the putative FM-QCP. The observed time-field scaling of
the muon asymmetry points to a cooperative mechanism and to the
presence of critical spin fluctuations. The overall behavior can not
be ascribed to either canonical spin glasses or to other
disorder-driven mechanisms.
The physics of CeFePO is different from other candidate systems where the putative FM-QCP is avoided by
a first order phase transition or a transition into an AFM ordered state.

The samples were synthesized by means of a two-step Sn-flux method.
The small crystals were powdered and pressed into pellets which are
referred to as polycrystalline (PC) sample in the subsequent
discussion. The large crystals were oriented and glued together with
silver paint to form a larger ``single crystal'' (SC). X-ray powder
diffraction confirms the ZrCuSiAs-structure type~\cite{Jesche2011}.
The synthesis conditions were different from the samples
investigated previously~\cite{Bruening2008,Jesche2011}.
However, the samples PC and SC as well as the sample of Ref.~\cite{Bruening2008}
are indistinguishable within the resolution of energy dispersive X-ray (EDX) measurements which confirm
the stoichiometric ratio 1:1:1:1.
Low temperature $\chi^\prime(T,B)$, $C(T,B)$
and $\mu$SR$(T,B)$ were measured in $^{3}$He-$^{4}$He-dilution
refrigerators. A commercial SQUID-VSM (Quantum Design) was used to
measure $\chi^\prime$ above $1.8\,$K. The $\mu$SR experiments were
performed on the $\pi$M3 beam line at the Swiss Muon Source at the
Paul-Scherrer-Institut, Switzerland. The $\mu$SR measurements were
executed in zero magnetic field and in applied magnetic fields
up to $0.75\,$T parallel to the initial muon-spin polarization (LF-$\mu$SR).

The first evidence of spin freezing is seen in the $T$ dependence of
$\chi^\prime(T)$ for the PC sample (Fig.~\ref{fig1}a). At $B =
0$ a distinct peak is found at $T_{g} = 0.67(1)$\,K where the
susceptibility reaches values as high as $2.8 \cdot
10^{-6}$\,m$^{3}$/mol. The $\chi^\prime(T)$ value at $2\,$K is three times
larger than the one measured in the sample of
Ref.~\cite{Bruening2008}. With increasing field its amplitude
decreases and $T_{g}$ shifts slightly and, above $0.8$\,T,
$\chi^\prime(T)$ flattens.
\begin{figure}[t]
\begin{center}
\includegraphics[width=0.95\columnwidth]{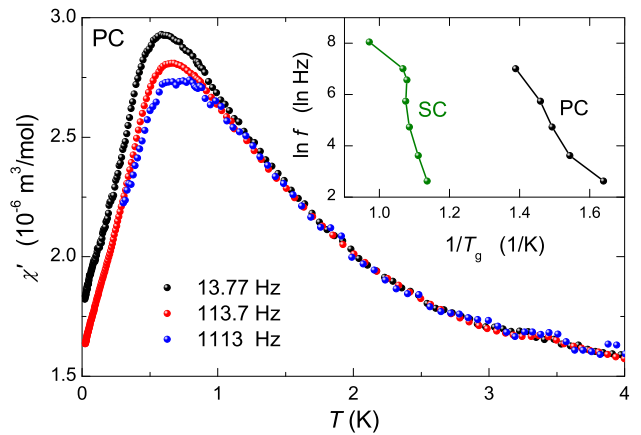}
\end{center}
\caption[1]{(Color online) Frequency dependence of $\chi^\prime(T)$
for the PC sample. Inset: Arrhenius plot for the PC sample (black)
and for the SC sample (green) with $B \perp c$.} \label{fig2}
\end{figure}
The small hump between $3$ and $9\,$K is only seen at $B = 0$ in the PC sample.
It is most likely due to an impurity phase, as it is completely suppressed in a small field of $0.02\,$T while $T_{g}$ does not change.
Moreover, a comparable feature is absent in the SC case.
Dissipative effects are corroborated by a peak in
$\chi^{\prime\prime}(T)$ at about $0.45\,$K (inset of
Fig.~\ref{fig1}a), which is a significantly lower temperature than
that of the maximum in $\chi^\prime(T)$. The same effects and $B$
dependence are observed in the SC sample, but at a higher $T_{g} = 0.92(2)\,$K.
This could be due to a very tiny difference in
stoichiometry, since both the PC and SC samples were taken from small and
large crystals, respectively, of the same batch which likely form at
different times during the growth.
The susceptibility is very
anisotropic ($\chi_{\perp c}/\chi_{\| c} \approx 5$), and with $B
\perp c$ it reaches a high peak value of $9 \cdot
10^{-6}\,$m$^{3}$/mol (Fig.~\ref{fig1}b and c) which is much larger than in the PC case.
This observation confirms the presence of anisotropic FM spin fluctuations, which are
much stronger along the basal planes. To check the bulk nature of
the freezing, we have measured the specific heat of the PC sample
(Fig.~\ref{fig1}d). A broad maximum emerges at about 0.55\,K in a
$C/T$ vs $T$ plot. While at $0.1\,$T the maximum is unchanged,
larger fields suppress it. The entropy difference between the
zero-field and the high-field curves is small, about
1\% of $R\ln(2)$, which is due to the Kondo screening of the Ce moments.
\begin{figure}[t]
\begin{center}
\includegraphics[width=0.95\columnwidth]{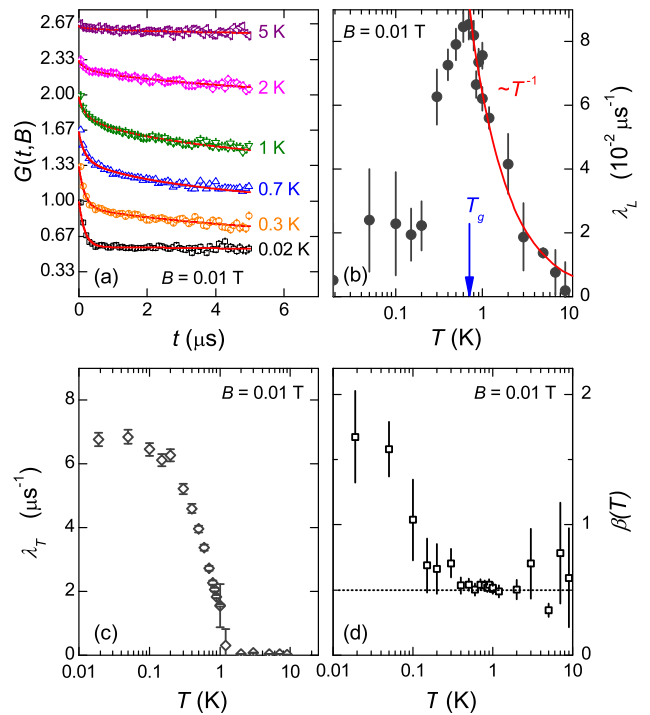}
\end{center}
\caption[1]{(Color online) (a) Normalized muon-spin asymmetry
function $G(t,B)$ at an applied field $B = 0.01\,$T for
representative temperatures above and below $T_{g}$. Solid lines are
fits according to Eq.~\ref{eq1}.
For clarity, the different curves are shifted subsequently by $0.33$.
(b) $T$ dependence of the dynamic $\mu$SR rate $\lambda_{L}$. The blue arrow marks the magnetic transition from dynamic to static magnetism at $T_{g} = 0.70(3)$\,K.
(c) $T$ dependence of the static $\mu$SR rate $\lambda_{T}$.
(d) $T$ dependence of the exponent $\beta$ in Eq.~\ref{eq1}. The dotted line denotes the value $\beta = 0.5$.
} \label{fig3}
\end{figure}
The data of the polycrystalline sample investigated in
Ref.~\onlinecite{Bruening2008} (gray data points in
Fig.~\ref{fig1}d) resemble the $C/T$ behavior at high temperatures,
while below 3\,K, $C/T$ remains constant without any indication of
freezing.

We have performed frequency ($f$) dependent measurements at
$B = 0$ on the PC and SC samples to investigate the spin
dynamics (Fig.~\ref{fig2}). Similar to spin glasses, the maximum in
$\chi^\prime(T)$ shifts to higher temperatures as the excitation
frequency is increased, while its amplitude decreases.
$T_{g}$ and $f$ are shown in an Arrhenius plot in the inset of Fig.~\ref{fig2}, from which the frequency shift $\delta$~\cite{Mydosh1993} can be evaluated.
For the PC and the SC sample we obtain
$\delta_{\mathrm{PC}}=0.085(11)$ and
$\delta_{\mathrm{SC}}=0.065(16)$, respectively. These values are
larger than those found for canonical spin glasses ($\delta \approx
0.005 ... 0.06$), yet they are below typical values of
superparamagnets ($\delta \approx 0.3$)~\cite{Mydosh1993}.
\begin{figure}[t]
\begin{center}
\includegraphics[width=0.95\columnwidth]{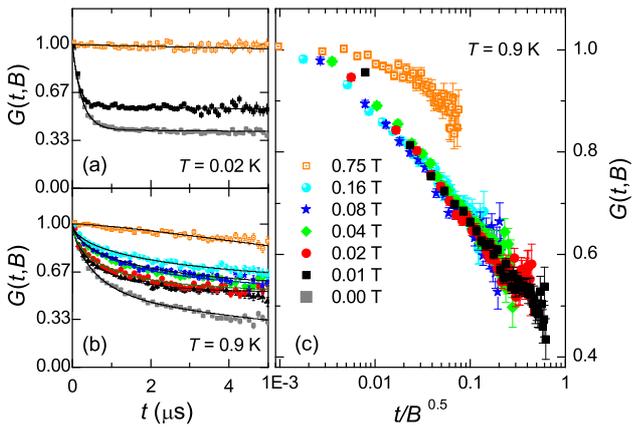}
\end{center}
\caption[1]{(Color online)
(a) Field dependence of the muon-spin asymmetry function $G(t,B)$ in CeFePO at $T=0.02\,$K
and (b) $T=0.9\,$K.
(c) The same data as in (b), plotted as a function of the scaling variable $t/B^{0.5}$.
} \label{fig4}
\end{figure}
A qualitatively similar behavior was found for
CePd$_{\mathrm{1-x}}$Rh$_{\mathrm{x}}$, where $\delta$ increases
when approaching the critical point at $x =
0.87$~\cite{Westerkamp2009}. It is interesting to note that in
CeFePO the sample with lower $T_{g}$ exhibits larger
$\delta$, too. This suggests that the QGP scenario might be also
applicable in CeFePO. On the other hand, the magnetic anisotropy and
the lack of evidence for FM cluster formation in zero-field-cooled/field-cooled
magnetization measurements (not shown) rule out such a mechanism in
CeFePO. Moreover, we could not fit our data with $\chi\propto C/T
\propto T^{\mathrm{\lambda-1}}$ with $0\leq\lambda\leq1$ in any
reasonable $T$-range~\cite{CastroNeto1998}.

To study the microscopic nature of the low-$T$ magnetism in CeFePO,
$\mu$SR measurements in zero and small longitudinal fields were
performed on the PC sample. $\mu$SR in small fields is dominated by
thermally excited Ce-4$f$ electronic spin fluctuations that couple
to the implanted muons. Figure~\ref{fig3}a displays the temperature
evolution of the normalized muon-spin asymmetry function $G(t,B)$ at a constant
field of $B=0.01\,$T which is sufficient to quench the weak
static relaxation due to nuclear dipole fields at the muon sites,
leaving only the dynamic and static contributions due to the
electronic Ce-4$f$ magnetic moments. The quantitative analysis takes
into account both static and dynamical fields: The static relaxation
dominates $G(t,B)$ at short times $t$, while at long $t$ the
relaxation rate probes only the dynamic spin fluctuations, i.e., the
Fourier transform (FT) of the dynamic spin-spin autocorrelation
function $q(t)=\langle
\textbf{S}_{i}(t)\cdot\textbf{S}_{i}(0)\rangle$. To account for both
contributions we use the following fitting function
\begin{equation}
\label{eq1}
G(t,B)=G_{1}\exp[-(\lambda_{T} t)]+G_{2}\exp[-(\lambda_{L} t)^{\beta}],
\end{equation}
with static (transversal) and dynamic (longitudinal) relaxation
rates $\lambda_{T}$ and $\lambda_{L}$, respectively. The fits
provide a very good description of the experimental data (solid
lines in Fig.~\ref{fig3}a). At low $T$ the spectra show nearly no
muon-spin relaxation at long times, i.e., very small relaxation
rates $\lambda_{L}$. Upon increasing $T$, $\lambda_{L}$ increases and
reaches a maximum at $T_{g} = 0.70(3)\,$K (Fig.~\ref{fig3}b) in
agreement with $T_{g}$ found in $\chi^\prime(T)$. Subsequently, it
decays following a $T^{-1}$ behavior up to 10\,K after which no
dynamic relaxation is observed as expected from/in accordance NMR
and susceptibility experiments~\cite{Bruening2008}. On the contrary,
the static component $\lambda_{T}$ increases steeply below $T_{g}$
up to a value of 7\,$\mu$s$^{-1}$ (Fig.~\ref{fig3}c). Its behavior resembles that of
the magnetic order parameter when entering a magnetically ordered
phase. The $\beta$ value of about 0.5 for $T \geq T_{g}$ (Fig.~\ref{fig3}d) indicates a broad inhomogeneous distribution of
fluctuating dynamical local fields (or relaxation rates)~\cite{MacLaughlin2004}. In the
ordered phase $\beta$ increases, reaching a value of about 1.7 at
$T=0.02\,$K, indicating that the spin fluctuations become static.
Figure \ref{fig4}a displays $G(t,B)$ at $0.02\,$K. At
$B = 0$, the absence of a spontaneous muon-spin precession
frequency indicates short-range magnetic order.
Such strongly damped $\mu$SR signal allows for an estimation of the
magnetic coherence length $\xi <10 \cdot a$, where $a$ is the lattice constant~\cite{Yaouanc2011}.
The observation of a $2/3$ and $1/3$ signal fraction below $T_{g}$ proves that $100\,$\% of the sample volume shows static magnetic order~\cite{Comment02}.
The muon-spin relaxation is completely suppressed at $B =
0.75\,$T demonstrating that the internal field distribution is
static in nature at $T = 0.02\,$K. Increasing $T$, a dynamic
contribution to the muon-spin relaxation develops (Fig.~\ref{fig4}b).\\
Glassy spin dynamics generally result in long-time correlations with
distinct signatures when $T_{g}$ is approached from high
$T$~\cite{Keren1996}. Theoretically, $q(t)$ is predicted to exhibit
power-law $q(t)=c t^{-\alpha}$ or stretched exponential
$q(t)=c\exp[-(\Lambda t)^{K}]$ behavior at $T>T_{g}$, that in both
cases can lead to a characteristic time-field scaling
$G(t,B)=G(t/B^{\gamma})$ after Fourier transforming $q(t)$, where $\gamma<1$ and $\gamma>1$
for power-law and stretched exponential correlations,
respectively~~\cite{Keren1996}. If this equation is obeyed a plot of
$G(t,B)$ versus $t/B^{\gamma}$ at $T>T_{g}$ will be universal. Figure \ref{fig4}b displays ${G}(t,B)$ at $T=0.9$\,K, which
is slightly above $T_{g}$, both in zero field and in magnetic longitudinal field between
$0.01$ and $0.75\,$T. The observed $B$ dependence corresponds to a
measurement of the FT of $q(t)$ over the frequency range
$\gamma_{\mu}B/2\pi\approx$ $1.4-100\,$MHz, where
$\gamma_{\mu}=2\pi\times 135.53\,$MHz/T is the muon gyromagnetic
ratio. As shown, the relaxation slows with increasing $B$. For low
enough $B$, the $B$ dependence is expected to be due to the change
of $\gamma_{\mu}B$ rather than an effect of field on
$q(t)$~\cite{MacLaughlin2004}. A breakdown of time-field scaling is
expected for high fields where $q(t)$ is directly affected by the
applied fields. Figure \ref{fig4}c shows the same
muon-spin asymmetry data as a function of $t/B^{\gamma}$. For
$\gamma=0.5(1)$ the data scale well over nearly 2.5 orders of
magnitude in $t/B^{\gamma}$ for all applied fields except for
$0.75\,$T, as expected for fields with $\mu_{B}B\geq k_{B}T$, which
should affect $q(t)$. The obtained scaling exponent
$\gamma=0.5(1)<1$ implies that, within the $\mu$SR frequency range,
$q(t)$ is well approximated by a power law, suggesting
cooperative and critical spin fluctuations rather than a
distribution of local fluctuation rates~\cite{MacLaughlin2004}.
This is in contrast to $\beta = 0.5$ which indicates a broad inhomogeneous
distribution of local fluctuation rates.
The cooperative behavior is supported by the fact that CeFePO is a stoichiometric system and the narrow NMR linewidth proves that it is locally not disordered~\cite{Bruening2008}.
Short-range correlations set in below $10\,$K, broadening the linewidth, in agreement with the $\mu$SR results.
The value of $\gamma$ seems to be weakly $T$ dependent (it is $0.4$ at
about $2\,$K, not shown) which suggests slow quantum rather than
thermal fluctuations.
From the $B$ dependence of $\lambda_{L}$, the
spin autocorrelation time $\tau_{c}$ can be estimated using
$\lambda_{L}(B)=(2\gamma_{\mu}^{2}\langle
B_{\textrm{loc}}^{2}\rangle\tau_{c})/[1+(
\gamma_{\mu}^{2}B^{2}\tau_{c}^{2})^{p}]$~\cite{Adroja2008}. Here,
$B_{\textrm{loc}}(t)$ describes the time-varying local magnetic
field at the muon site due to fluctuations of neighboring Ce-4$f$
moments, with a local time averaged second moment
$\gamma_{\mu}\langle B_{\textrm{loc}}^{2}\rangle$. For
$\hbar\omega\gg k_{B}T$, the fluctuation-dissipation theorem relates
$\tau_{c}$ to the imaginary component of the local $q$-independent
$f$-electron dynamic susceptibility, i.e., $\tau_{c}(B) \approx
(k_{B}T /
\mu_{B}^2)\,(\chi^{\prime\prime}(\omega)/\omega)$~\cite{Toll1956}. A
fit to the data (not shown) yields
$\tau_{c}\approx1.8\times10^{-8}$\,s and $p=0.67$. The value for
$\tau_{c}$ indicates very slow or glass-type spin dynamics.

In conclusion, we have shown that single crystals of CeFePO, located close to a FM instability, show spin-glass-like freezing.
It is evidenced by the frequency-dependent peak in $\chi^\prime(T)$ at $T_{g}$, the broad maximum in $C(T)/T$ as well as clear signatures in $G(t,B)$, which clearly evidences the transition from dynamic to short-range static magnetism.
The frequency shift of $T_{g}$ suggests values slightly larger than for canonical spin glasses, indicating the presence of large fluctuating regions above $T_{g}$, and the time-field scaling strongly suggests cooperative behavior.
Our results imply that the putative FM-QCP is avoided in a new manner: We do not observe a first order phase transition or AFM order, but rather a transition into a short-range ordered state.
Moreover, the magnetic anisotropy, the lack of evidence for FM clusters and the time-field scaling rule out a disorder-driven scenario (e.g., the QGP) as the mechanism underlying the spin dynamics in CeFePO.
We might have in CeFePO a possible combination
of both scenarios: The close proximity of CeFePO to a FM instability and its magnetic anisotropy seem to drive the system to develop short-range magnetic correlations which might have their origin in the mechanism described in Refs.~\onlinecite{Kirkpatrick2012,Chubukov2004,Conduit2009}.
Below $T_{g}$, magnetic short-range order then forms with a certain texture (e.g., homogeneous cobbled magnetically ordered regions of different sizes) which would explain the spatially distributed $\mu$SR rates.

We are indebted to G. J. Conduit, A. G. Green, D. E. MacLaughlin and T. Vojta for useful discussions, and we acknowledge the PSI accelerator crew.
This research project has been supported by the European Commission under the 7th Framework Programme through the 'Research Infrastructures' action of the 'Capacities' Programme, NMI3-II Grant number 283883.
Part of this work has been supported by the DFG Research Unit 960 ``Quantum Phase Transitions''.


\begin{thebibliography}{10}
\bibitem{Comment02}
Please note, a constant $T$ independent background signal due to muons that
  stopped in a Ag sample holder (signal fraction $= 15\,$\%) was taken into
  account .
\end{thebibliography}
\end{document}